\documentclass[a4paper,12pt]{article}
\usepackage[latin1]{inputenc}
\usepackage{amsmath}
\usepackage{amsfonts}
\usepackage{amssymb}
\usepackage{graphicx}

\title{ Relativistic Potentials with Rational Extensions }
\author{K. Haritha and K. V. S. Shiv Chaitanya\footnote{chaitanya@hyderabad.bits-pilani.ac.in},\\
Department of Physics, BITS Pilani, Hyderabad Campus,  \\Jawahar Nagar, Kapra Mandal, Medchal Dist, Hyderabad,\\ Telangana India 500 078.}
\date{}
\begin{document}

\maketitle

\begin{abstract}
	In this paper, we construct  isospectral Hamiltonians without shape invariant potentials for the relativistic quantum mechanical potentials such as the Dirac Oscillator and Hydrogen-like atom. 
\end{abstract}
\section{Introduction}
The discovery of the new orthogonal polynomials, known as  exceptional polynomials, in the last decade by David G´Omez-Ullate \textit{et al} \cite{kam,kam1}, has given rise to a lot of new exactly solvable potentials. These new potentials are known as rational extensions. The old and the new potentials are related by an intertwining operator of the supersymmetric quantum mechanics. These new potentials give rise to a rich structure which is one of the states  missing between the old and the new Hamiltonians. This, in turn, is reflected in the solution as one of the degrees missing in the polynomial sequence. 
In literature, the missing degrees in the polynomial sequence are termed as Codimension of the exceptional family. It is well known in the literature that these rational extensions are related through Darboux transformations. First such potential of codimension-2 is constructed by Quesne  \cite{qu,qu1} and the potential for any arbitrary codimension is given by Odake \& Sasaki \cite{sak2,sak3}. Exceptional orthogonal families with higher order are generated through Darboux transformations \cite{sak,gkm1,gar1}. 
The role of Darboux transformations is further studied and  clarified in the reference \cite{gkm}.
Exceptional polynomials are developed  using the prepotential approach \cite{ho}.  The exceptional polynomials are also extended to \cal{PT}  symmetry Hamiltonians \cite{bachi} and the prominent examples are PT-symmetric Scarf II potential \cite{bach1, bach2, bach3}. Exceptional polynomials for symmetry group preserving the form of  Rayleigh-Schr\"odinger equation are constructed \cite{gar2}.  Alternative derivation  of many infinitely  exceptional Wilson and Askey-Wilson polynomials is presented \cite{od1}. These exceptional polynomial systems appear as solutions to the quantum mechanical problems  in the exactly solvable models \cite{ku,di} or in superintegrable systems \cite{mat}.  In the reference \cite{kvs}, it is shown that the quantum mechanical problems whose solutions are Laguerre or Jacobi polynomials will also admit exceptional Laguerre or Jacobi polynomial solutions after the suitable modification of the potential. One of the authors has constructed the exceptional isospectral Hamiltonians for the harmonic oscillator Wigner function \cite{kvs1}.
 Particles whose energies are greater than their rest mass energy, where the relativistic effects creep in, are to be invariably studied under relativistic quantum mechanics. The study of subatomic particles, moving at relativistic speeds, is incomplete without unifying quantum mechanics with relativity. 
The properties of the Dirac Oscillator potential are studied by J. Benitez  \textit{et al} \cite{ben}. It is found that the energy eigen functions are the same as the non relativistic harmonic oscillator\cite{rel}. In a work by \cite{can}, a canonical quantization was done on the Dirac Oscillator in (1+1) and (1+3) dimensions where it was found that Dirac oscillator field can be identified to be composed of infinite degrees of freedom that are decoupled quantum linear harmonic oscillators. It is shown that the energies of relativistic linear harmonic oscillators for the Dirac oscillator field are the quanta.
In a work done by \cite{dir}, the partition function and other thermal properties like free energy, entropy and specific heat are calculated for the (1+1) dimensional Dirac Oscillator with spin non-commutativity of coordinates.
In this paper, we construct the rational extensions of certain  relativistic quantum mechanical problems. The exceptional Dirac equation under Hartmann and Ring shaped potentials\cite{har}, the Dirac Oscillator problem under the Aharnov Bohm and Magnetic monopole potentials \cite{ahar}  the hydrogen-like atom are solved for super symmetric partner potentials, whose solutions turn out to be exceptional polynomials. \par
This paper is arranged as follows:
In section II, we consider the Dirac equation in the presence of Hartmann and ring shaped potentials  involving the Laguerre and Jacobi differential equations  to radial and angular parts respectively, where we derive the exceptional polynomial solutions to radial and angular parts. In section III, we take the Radial component of the Dirac oscillator wherein the electromagnetic potential is the combination of the Aharnov-Bohm and magnetic monopole potential and after making suitable modifications to the equation, we obtain the exceptional Laguerre solutions. In section IV, we derive  isospetral potential  the relativistic Hydrogen-like atom with rational extensions containing the exceptional Laguerre polynomial as their solutions. In section V, we make brief concluding remarks.
\section
{The exceptional polynomial solutions for the Radial Part of the Hartmann and ring shaped potentials}
In this section, we construct the  exceptional Laguerre polynomials as solutions to the Dirac equation in the presence of Hartmann and Ring shaped potentials.
Hartmann potential is a non-central potential introduced by Hartmann  \cite{hp} obtained by adding a potential to the coulomb potential. Ring shaped oscillator potential is that obtained when the coulomb part of the Hartmann potential is replaced by Harmonic Oscillator term used to describe its energy spectrum \cite{hp}. We consider the radial part of the Dirac equation in the presence of Hartmann and Ring shaped potentials that is derived by Z. Bakhshi \cite{har}.
\begin{equation}
\frac{d^2u(r)}{dr^2}+[\frac{(\epsilon^2-M^2c^4)}{\hbar^2c^2}-\frac{\rho}{r^2}+[\frac{(\epsilon+Mc^2)}{\hbar^2c^2}]\frac{V_o\lambda}{r}]u(r)=0 \label{1}
\end{equation}
By following the procedure in ref \cite{har} and defining 
$\rho=l(l+1); (\frac{(\epsilon+Mc^2)}{c^2}\frac{V_o\lambda}{r})=\frac{1}{4}\omega^2;
\frac{\epsilon^2-M^2c^4}{c^2}=2n\omega+(l+\frac{3}{2})\omega$ then 
substituting in the equation (\ref{1}) in terms of natural units we get
\begin{equation}
\frac{d^2u(r)}{dr^2}+[2n\omega+(l+\frac{3}{2})\omega-\frac{l(l+1)}{r^2}-\frac{1}{4}\omega^2r^2]u(r)=0. \label{jui}
\end{equation}
The solution to equation (\ref{jui})  as taken by Z. Bhakshi(\cite{har} )
\begin{equation}
U_{n,l}(r)= (\xi)^{\frac{l+1}{2}}exp(-\frac{\xi}{2})L_n(\xi),
\end{equation}
where $\xi=\frac{1}{2}\omega r^2$ gives  the Laguerre differential equation and $L_n(\xi)$ satisfies
\begin{equation}
\xi L_n''(\xi)+[(l+\frac{3}{2}-\xi)]L_n'(\xi)+n L_n(\xi)=0.\label{lauh}
\end{equation}
To derive exceptional Laguerre polynomials as solutions to the Dirac equation in the presence of Hartmann and Ring shaped potentials, we add an extra potential $V_e$ to the differential equation (\ref{lauh}).
\begin{equation}
\xi H_n''(\xi)+[(l+\frac{3}{2}-\xi)]H_n'(\xi)+(n+V_e) H_n(\xi)=0, \label{eqn4}
\end{equation}
here we define $l+\frac{1}{2}=m $ and $n=\lambda-1$. We demand that
 \begin{equation}
 H(\xi)=\frac{f_n(\xi)}{\xi+m} 
\end{equation}
be the solution to the equation(\ref{eqn4}). Then we get
\begin{eqnarray}
\\&&\nonumber
\xi(\frac{2f(\xi)}{(\xi+m)^3}-\frac{2f'(\xi)}{(\xi+m)^2}\\&&\nonumber +\frac{f''(\xi)}{\xi+m})+(m+1-\xi)(\frac{-f(\xi)}{(\xi+m)^2}\\&&\nonumber
+\frac{f'(\xi)}{\xi+m})+(\lambda-1+V_e(\xi,m))\frac{f_n(\xi)}{\xi+m}=0
\end{eqnarray}
\begin{eqnarray}
\xi{f''(\xi)}-[\frac{\xi-m}{\xi+m}](m+1+\xi)f'(\xi)+\frac{2m}{(\xi+m)^2}-\frac{1}{\xi+m}+V_e)f(\xi)=-(\lambda-1)f(\xi)
\end{eqnarray}
On rearranging
\begin{eqnarray}
\xi{f''(\xi)}+[\frac{-2\xi}{\xi+m}+(m+1-\xi)f'(\xi)+(\frac{2\xi}{(\xi+m)^2}-\frac{m+1-\xi}{\xi+m}+V_e)f(\xi)=-(\lambda-1)f(\xi)
\end{eqnarray}
\begin{eqnarray}
-\xi{f''(\xi)}+[\frac{\xi-m}{\xi+m}][(m+1+\xi)f'(\xi)+\frac{2m}{(\xi+m)^2}-\frac{1}{\xi+m}+V_e]f(\xi)=(\lambda-1)f(\xi) \label{eqn1}
\end{eqnarray}
On comparing the equation (\ref{eqn1})  with the standard form of $X_1$- exceptional Laguerre equation in eqn.(\ref{lag}),
\begin{equation}
-xy''+\frac{x-k}{x+k}[(k+x+1)y'+y]=\lambda(y) \label{lag}
\end{equation}
and demanding  $f_n(\xi)$ satisfy the $X_1$- exceptional Laguerre equation determines the potential $V_e$ as
$V_e(\xi,m)$ to be 
\begin{equation}
V_e(\xi,m)= -\frac{2m}{(\xi+m)^2}+\frac{1}{\xi+m}.
\end{equation}
Substituting back $m= l+\frac{1}{2}$ and $\xi=\frac{1}{2}\omega r^2$
we obtain the exceptional partner potential as
\begin{equation}
V_e(r,l)= -\frac{2l+1}{(\frac{1}{2}\omega r^2+(\frac{2l+1}{2}))^2}+\frac{1}{(\frac{1}{2}\omega r^2+\frac{2l+1}{2})}
\end{equation}
Thus, we have to add these terms to the original potential in order to get the exceptional partner potential
$V^+(r,l)=V_o(r,l)+V_e(r,l)$.
 Hence, we obtain the modified form of the partner potential as
\begin{equation}
V^+(r,l)= \frac{(\epsilon^2-M^2c^4)}{\hbar^2c^2}-\frac{\rho}{r^2}+[\frac{(\epsilon+Mc^2)}{\hbar^2c^2}]\frac{V_o\lambda}{r}-\frac{2l+1}{(\frac{1}{2}\omega r^2+(\frac{2l+1}{2}))^2}+\frac{1}{(\frac{1}{2}\omega r^2+\frac{2l+1}{2})}
\end{equation}
Thus by changing the weight function of the Laguerre polynomial, we arrive at an isospectral potential to the Hartmann and ring shaped potentials.
\section*{The Exceptional polynomial solutions for the angular part of the Hartmann and ring shaped potentials}
In this section,  we shall proceed to consider the angular part of the ring shaped potentials as given by the first function in the differential equation from the paper \cite{har}
\begin{equation}
\frac{d^2H(\theta)}{d\theta^2}+[-\frac{m^2-\frac{1}{4}}{sin^2\theta}-(\epsilon+Mc^2)f(\theta)+\rho+\frac{1}{4}]H(\theta)=0
\end{equation}
substituting $f(\theta)=\frac{\gamma+\beta \cos\theta+\alpha \cos^2\theta}{sin^2\theta}$ as done in ref \cite{har}.
\begin{equation}
\frac{d^2H(\theta)}{d\theta^2}+[[-m^2-\frac{1}{4}-\eta(\gamma+\alpha)]\csc^2\theta-\eta\beta\csc\theta\cot\theta+\eta\alpha+\rho+\frac{1}{4}]
H(\theta)]=0
\end{equation}
Taking $\eta = \epsilon+Mc^2; 
\eta(\gamma+\alpha)+m^2-\frac{1}{4} = \lambda^2+s^2+s;
\eta\alpha+\rho+\frac{1}{4}= (s+n)^2;
\eta\beta= -\lambda(2s-1)$
\begin{equation}
\frac{d^2H(\theta)}{d\theta^2}+[-(\lambda^2+s^2+s)\csc^2\theta+\lambda(2s-1)\csc \theta \cot \theta +(s+n)^2]H(\theta)=0 \label{eqn2}
\end{equation}
Taking the solution for the equation(\ref{eqn2}) to be
\begin{equation}
H(\theta)= (1-\cos\theta)^{\frac{s-\lambda}{2}}(1+\cos\theta)^{\frac{s+\lambda}{2}} P_n^{(-\lambda+s-\frac{1}{2}, \lambda+s-\frac{1}{2})}(\cos\theta)
\end{equation}
We obtain
\begin{eqnarray}
(1-cos^2\theta)P_n''(cos\theta)
+[2(\lambda-scos\theta)-cos\theta]P_n'(cos\theta)
+(n^2+2sn)P_n(cos\theta)=0 \label{new}
\end{eqnarray}
\label{eqn6}
Here $\alpha=-\lambda+s-\frac{1}{2}$ ;
$\beta=\lambda+s-\frac{1}{2}$
where $P_n^{(\alpha, \beta)}(cos\theta)$ represents the Jacobi polynomial. On suitable modification of the weight function  of the Jacobi equation, we can get the extra terms of the exceptional potential.\\
The exceptional Jacobi differential equation is given by
\begin{equation}
(z^2-1)f_n''(z)+\frac{2a(1-bz)}{b-z}[(z-c)f_n'(z)-f_n(z)]=0 \label{why}
\end{equation}
Now adding an extra potential term, $V_e$ to the equation(\ref{new}) we have,
\begin{equation}
(1-cos^2\theta)H_n^{''(\alpha, \beta)}(cos\theta)
+[2(\lambda-scos\theta)-cos\theta]H_n^{'(\alpha, \beta)}(cos\theta)
+[(n^2+2sn)+V_e]H_n^{(\alpha, \beta)}(cos\theta)=0 \label {problem}
\end{equation}
Here 
$ a=\frac{1}{2}(\beta-\alpha) = \lambda ; b=\frac{\beta+\alpha}{\beta-\alpha}=\frac{2s-1}{2\lambda}$; 
$ c=b+\frac{1}{a}=\frac{2s+1}{2\lambda}$
We demand
\begin{equation} H_n^{(\alpha, \beta)} (cos\theta) =\frac{f_n^{(\alpha, \beta)} (cos\theta)}{\frac{2s-1}{2\lambda}-cos\theta}
\end{equation}
to be the solution of the equation(\ref{problem})\\
(The calculations of $\frac{dH_n^{(\alpha, \beta)}}{dcos\theta} and \frac{d^2H_n^{(\alpha, \beta)}}{dcos\theta^2}$ 
are provided in the Appendix1.)\\
Substituting the values in the equation(\ref{problem})
\begin{eqnarray}
\\&&\nonumber (cos^2\theta-1)\frac{df_n(cos\theta)}{d(cos\theta)}\\&&\nonumber +[2\lambda-(2s+1)cos\theta-\frac{2sin^2\theta}{cos\theta-\frac{2s-1}{2\lambda}}]\frac{d^2f_n(cos\theta)}{d(cos\theta)^2}\\&&\nonumber
+[n(n+2s)+\frac{2sin^2\theta}{(cos\theta-\frac{2s-1}{2\lambda})^2}
+\frac{2\lambda-2scos\theta-cos\theta}{cos\theta-\frac{2s-1}{2\lambda}}
-\frac{cos\theta}{cos\theta-\frac{2s-1}{2\lambda}}]f_n(cos\theta)= 0
\end{eqnarray}
\begin{eqnarray}
\\&&\nonumber (cos^2\theta-1) \frac{df_n}{d(cos\theta)}\\&&\nonumber 
+[\frac{2\lambda(1-(\frac{2s-1}{2\lambda})cos\theta)cos\theta}{\frac{2s-1}{2\lambda}-cos\theta}-\frac{2\lambda(1-(\frac{2s-1}{2\lambda})cos\theta)\frac{2s+1}{2\lambda}}{\frac{2s-1}{2\lambda}-cos\theta}]\frac{d^2f_n}{d(cos\theta)^2}
\\&&\nonumber
+[n^2+2sn+\frac{2sin^2\theta}{(\frac{2s-1}{2\lambda}-cos\theta)^2}+\frac{2\lambda(1-(\frac{2s-1}{2\lambda})cos\theta)}{\frac{2s-1}{2\lambda}-cos\theta}-\frac{2cos\theta}{\frac{2s-1}{2\lambda}-cos\theta}]f_n(cos\theta) = 0
\end{eqnarray}
The derivation of the Exceptional form of Jacobi differential equation is provided in the Appendix2.
taking $cos\theta = z; a=\frac{\beta-\alpha}{2}=\lambda; b=\frac{\beta+\alpha}{\beta-\alpha}=\frac{2s-1}{2\lambda}; c=b+\frac{1}{a} =\frac{2s+1}{\lambda}$\\
This is of the form of exceptional Jacobi differential equation given by \ref{why} comparing with the standard form we get the extra terms of the exceptional potential$ V_e$ this determines the exceptional potential term $V_e(cos\theta)$to be
\begin{equation}
V_e(cos\theta)=-[\frac{2(1-cos^2\theta)}{(\frac{2s-1}{2\lambda}-cos\theta)^2}-\frac{2cos\theta}{\frac{2s-1}{2\lambda}-cos\theta}]
\end{equation}
After a few algebraic modifications we get
\begin{equation}
V_e(cos\theta)=\frac{2(\frac{2s-1}{2\lambda})}{(\frac{2s-1}{2\lambda}-cos\theta)}-\frac{2-2(\frac{2s-1}{2\lambda})^2}{(\frac{2s-1}{2\lambda}-cos\theta)^2}
\end{equation}
Hence the total superpotential is given by $V(cos\theta)=V_o(cos\theta)+V_e(cos\theta)$
\begin{equation}
V(cos\theta)=-(\lambda^2+s^2+s)\csc^2\theta+\lambda(2s-1)\csc \theta \cot \theta +(s+n)^2-\frac{2-2(\frac{2s-1}{2\lambda})^2}{(\frac{2s-1}{2\lambda}-cos\theta)^2}+\frac{2(\frac{2s-1}{2\lambda})}{\frac{2s-1}{2\lambda}-cos\theta}
\end{equation}
Similarly we can find the exceptional Jacobi Solutions to the second function of the  reference paper\cite{har}; given equation
\begin{equation}
\frac{d^2H(r)}{dr^2}+[\frac{-\lambda(\lambda-1)}{sin^2\theta}-\frac{s(s-1)}{cos^2\theta}+(\lambda+s+2n^2)]H_n(cos\theta)
\end{equation}
The solution satisfies the equation 
\begin{equation}
H(\theta)= (1-cos2\theta)^{\frac{\lambda}{2}}(1+cos2\theta)^{\frac{s}{2}}f_n^{(\lambda-\frac{1}{2},s-\frac{1}{2})}(cos\theta).
\end{equation}
Now adding an extra potential term $V_e(cos\theta)$ to the above equation gives
\begin{equation}
\frac{d^2H(r)}{dr^2}+[\frac{-\lambda(\lambda-1)}{sin^2\theta}-\frac{s(s-1)}{cos^2\theta}+(\lambda+s+2n^2 +V_e(cos\theta))]H_n(cos\theta)
\end{equation}
$a= \frac{s-\lambda}{2}$ ;$b=\frac{s+\lambda-1}{s-\lambda}$ ;
$c=\frac{s+\lambda+1}{s-\lambda}$\\,
now setting
\begin{equation}
 H(cos\theta)=\frac{f(cos\theta)}{cos\theta-b}
 \end{equation}
and demanding that $f(cos\theta)$ satisfies the exceptional Jacobi differential equation
\begin{equation}
H(\theta)= (1-cos2\theta)^{\frac{\lambda}{2}}(1+cos2\theta)^{\frac{s}{2}}\frac{f_n^{(\lambda-\frac{1}{2},s-\frac{1}{2})}(cos\theta)}{\frac{s+\lambda-1}{s-\lambda}}
\end{equation}
Solving on the similar lines, the exceptional potential for the exceptional Jacobi differential equation, can be shown as
\begin{equation}
V_e(cos\theta)=\frac{2(\frac{s+\lambda-1}{s-\lambda})}{cos\theta-(\frac{s+\lambda-1}{s-\lambda})}-\frac{2-2(\frac{s+\lambda-1}{s-\lambda})^2}{(cos\theta-\frac{s+\lambda-1}{s-\lambda})^2}
\end{equation}
The complete isospectral potential is given by 
\begin{equation}
V(cos\theta)=\frac{-\lambda(\lambda-1)}{sin^2\theta}-\frac{s(s-1)}{cos^2\theta}+(\lambda+s+2n^2)+\frac{2}{cos\theta-(\frac{s+\lambda-1}{s-\lambda})}-\frac{2-2(\frac{s+\lambda-1}{s-\lambda})^2}{(cos\theta-\frac{s+\lambda-1}{s-\lambda})^2}
\end{equation}
\section{Exceptional polynomial solutions for the Dirac oscillator in the presence of Aharnov Bohm and magnetic monopole potentials}
The Aharnov Bohm effect is a quantum mechanical phenomenon in which charged particle, though confined to a region of no  electric and magnetic fields, is influenced by an electromagnetic potential. The three dimensional Dirac equation in the presence of Aharnov Bohm and magnetic monopole potentials is studied by A.H. Alhaidhari and the exact solutions to a Dirac Oscillator is obtained, vanishing the time component in the electromagnetic potential \cite {ahar}.
The three dimensional Dirac equation is given by
\begin{equation}
(i\gamma^\mu\partial_\mu-m)\psi=0
\end{equation}
where $m$ is the mass of rest particle. Here we consider natural units and the radial part of the dirac oscillator for positive energies is given by 
\begin{equation}
\left[-\frac{d^2}{dr ^2} + \frac{ \epsilon_\theta(\epsilon_\theta+1)}{r^2}+W^2
-\frac{dW}{dr}+2\epsilon_\theta\frac{W}{r}-\frac{\epsilon^2-1}{\lambda^2} \right] \tilde{R}_+(r) = 0. 
\label{relrad}
\end{equation}
For the Dirac-Oscillator problem, we take $W(r)$ to be linear in the
radial coordinate as $W =\omega^2r$, where the real parameter $\omega$ is the oscillator frequency. Thus we have
\begin{equation}
\left[-\frac{d^2}{dr ^2} + \frac{ \epsilon_\theta(\epsilon_\theta+1)}{r^2}+\omega^4 r^2
+\omega^2(2\epsilon _\theta-1)-\frac{\epsilon^2-1}{\lambda^2} \right] \tilde{R}_+(r) = 0. 
\label{relrad}
\end{equation}

By choosing $\epsilon_\theta=l$ and $E=\omega^2(2\epsilon_\theta-1)-\frac{\epsilon^2-1}{\lambda^2}$. Going over to the natural units, that is  $\omega=1$, the equation (\ref{relrad}) resembles non-relativistic 3D oscillator. Taking $R(r)= r^{l+1}exp(-r^2/2)$ as solution for the differential equation (\ref{relrad}), we have
\begin{equation}
f''(r)+[\frac{2(l+1)}{r}-2r]f'(r)+(-2l-3+2E)f(r)=0
\end{equation}
 By making a change of  variable
$r^2= \xi $, we arrive at the Laguerre form of differential equation,
\begin{equation}
\xi\frac{d^2}{d\xi^2}L_n^k(\xi)+(l+\frac{3}{2}-\xi)\frac{d}{d\xi}L_n^k(\xi)+\frac{1}{4}[2E-2l-3]L_n^k(\xi)=0
\end{equation}
where $L_n^k(\xi)$ satisfies Laguerre  differential equation.
Then by adding an extra potential $V_e^{osc}$ we have
\begin{equation}
\xi\frac{d^2}{d\xi^2}H(\xi)+(l+\frac{3}{2}-\xi)\frac{d}{d\xi}H(\xi)+\frac{1}{4}(2E-3-2l+V_e^{osc})H(\xi)=0
\end{equation}

Taking $m = l+\frac{1}{2} ; \lambda = 2E-2l-3$
$H(\xi)$ satisfies the exceptional Laguerre differential equation. On modification of the weight function and taking $\lambda=n-1$
\begin{equation}
H(\xi)= \frac{f^k(\xi)}{\xi+m} 
\end{equation}

\begin{equation}
\xi \frac{d^2}{d\xi^2}\frac{f^k(\xi)}{\xi+m}+(m+1-\xi)\frac{d}{d\xi}\frac{f^k(\xi)}{\xi+m}+(n-1)\frac{f^k(\xi)}{\xi+m}=0
\end{equation}
\begin{equation}
\xi\frac{d^2}{d\xi^2}f^k(\xi)-\frac{\xi-m}{\xi+m}(m+\xi+1)\frac{d}{d\xi}f^k(\xi)+\frac{\xi-m}{\xi+m}f^k(\xi) = -(n-1)f^k(\xi)
\end{equation}
on simplification we get
\begin{equation}
-\xi\frac{d^2}{d\xi^2}f^k(\xi)+\frac{\xi-m}{\xi+m}[(m+\xi+1)\frac{d}{d\xi}f^k(\xi)-f^k(\xi)]=(n-1)f^k(\xi)
\end{equation}
gives $V(\xi)$ to be
\begin{equation}
H(\xi)=\frac{\xi^{\frac{1}{2}}exp(-l/2)}{(\xi+\frac{2l+1}{2})}f^k(\xi)
\end{equation}
\begin{equation}
V_e(\xi,m)=\frac{1}{\xi+m}-\frac{2m}{(\xi+m)^2}
\end{equation}
\begin{equation}
V_e^(\xi,l)= \frac{1}{\xi+\frac{2l+1}{2}}-2\frac{\frac{2l+1}{2}}{(\xi+\frac{2l+1}{2})^2}
\end{equation}
Thus, we obtain the modified Dirac Oscillator potential in natural units, $V(r,l)=V_o(r,l)+V_e(r,l)$
\begin{equation}
V(r,l)= \frac{r^2}{2}+\frac{l(l+1)}{r^2}-E+\frac{1}{r^2+\frac{2l+1}{2}}-\frac{2l+1}{(r^2+\frac{2l+1}{2})^2}
\end{equation}
The potential having the same eigen values, obtained for the Dirac Oscillator under Aharnov Bohm and Magnetic monopole potentials is thus determined.
\section{Relativistic Hydrogen-like Atom}
The standard Hydrogen atom problem can be exactly solved using relativistic quantum mechanics. The radial wave function and its solution are derived which involve Associated Laguerre differential equation\cite{relhyd}. In this work we derive the Hamiltonian for relativistic  Hydrogen atom involving exceptional differential equation having the same energy eigen values as the classical Associated Laguerre differential equation.
The radial equation for the relativistic Hydrogen-like atom is given by
\begin{equation}
\left[\frac{d^2}{dr ^2} +\frac{2}{r}\frac{d}{dr} + \frac{E^2-m^2c^4}{\hbar^2c^2} + \frac{2EZe^2}{\hbar^2c^2r}
 - \frac{ l(l+1)-z^2e^4/\hbar^2c^4}{r ^2} \right] R(r) = 0. 
\label{rad}
\end{equation}
By defining the following
$l(l+1)-z^2e^4/\hbar^2c^4=s(s+1);
\frac{2Eze^2}{\hbar^2c^2}=\lambda ; and 
\frac{E^2-m^2c^4}{\hbar^2c^2}=\chi^2/4$
Then making a change of variable $y=\chi r$, where $\chi=\frac{2}{\hbar c}(m^2c^4-E^2)^\frac{1}{2}$ 
\begin{equation}
\frac{1}{r}\frac{d^2}{dr^2}rR(r)+(\frac{\lambda}{r}+\frac{\chi^2}{4}-\frac{s(s+1)}{r^2})
R(r)=0
\end{equation}
By taking $rR(r)=U(r)$
\begin{equation}
\frac{d^2}{dr^2}U(r)+(\frac{\lambda}{r}+\frac{\chi^2}{4}-\frac{s(s+1)}{r^2})
U(r)=0
\end{equation}
The solution is given by
$U(r)=r^{s+1}exp(-r/2)L_n^k(r)$\\
$\frac{d^2U}{dr^2}=r^{s+1}exp(-r/2)[L_n^{k"}(r)+[\frac{2(s+1)}{r}-1]L_n^{k'}(r)+(\frac{1}{4}-\frac{s+1}{r}+\frac{s(s+1)}{r^2})L_n^k(r)]$\\
\begin{equation}
r^{s+1}exp(\frac{-r}{2})[\frac{d^2}{dr^2}L_n^k(r)+(\frac{2(s+1)}{r}-1)\frac{d}{dr}L_n^k(r)+(\frac{1}{4}-\frac{(s+1)}{r}+\frac{s(s+1)}{r^2}+\frac{\chi^2}{4}+\frac{\lambda}{r}-\frac{s(s+1)}{r^2})L_n^k(r)]=0
\end{equation}\label{colp}
  where $L_n^k(r)$ satisfies Laguerre differential equation.
  Now, adding an extra potential, say $V_e(r,s)$ to the equation(\ref{colp}), we have
\begin{equation}
  r\frac{d^2}{dr^2}H_n(r) + ((2s+2-r) \frac{d}{dr}H_n(r) +(\lambda -s-1+V_e(r,s))H_n(r) = 0.\label{colpr}
\end{equation} 
We shall modify the weight function to obtain the exceptional potentials.\\
$H_n(r)= \frac{f(r)}{r+2s+1};
\frac{dH_n}{dr}=\frac{-f(r)}{(r+2s+1)^2}+\frac{f'(r)}{r+2s+1} ;
\frac{d^2H_n}{dr^2}=f''(r)-\frac{2f'(r)}{(r+2s+1)^2}-\frac{2f(r)}{(r+2s+1)^3} \label{eqn7} $\\
Substituting in the equation (\ref{colpr})
\begin{equation}
r\frac{d^2}{dr^2}(\frac{f(r)}{r+2s+1})+(2(s+1)-r)\frac{d}{dr}(\frac{f(r)}{r+2s+1})+(n-1)\frac{f(r)}{r+2s+1}=0.
\end{equation}
On simplification,
\begin{eqnarray}&=&
r[\frac{f''(r)}{r+2s+1}-\frac{2f'(r)}{(r+2s+1)^2}
+\frac{2f(r)}{(r+2s+1)^3}]\\&&\nonumber
+(2s+2-r)[-\frac{f(r)}{(r+2s+1)^2}+\frac{f'(r)}{(r+2s+1)}]\\&&\nonumber
+[\lambda-s-1+V_e(r,s)]\frac{f(r)}{r+2s+1}
\end{eqnarray}

\begin{equation}
rf''(r)+[\frac{-2r}{r+2s+1}+(2s+2-r)]f'(r)+[\frac{2r}{(r+2s+1)^2}-\frac{2s+2-r}{r+2s+1}+(\lambda-s-1+V_e(r,s))]f(r)=0
\end{equation}
\begin{equation}
rf''(r)+[(\frac{r-2s-1}{r+2s+1})(2s+2-r)]f'(r)+[\frac{r-2s-1}{r+2s+1}+\frac{r-2s-1}{(r+2s+1)^2}+(\lambda-s-1+V_e(r,s))]f(r)=0
\end{equation}
We assume $\lambda-s=n$
\begin{equation}
rf''(r)-(\frac{r-2s-1}{r+2s+1})[(2s+2-r)f'(r)+\frac{2(2s+1)}{(r+2s+1)^2}-\frac{1}{r+2s+1}+V_e(r,s))]f(r)=-(n-1)f(r)
\end{equation}
Writing in the form of Laguerre differential equation 
\begin{equation}
-rf''(r)+(\frac{r-2s-1}{r+2s+1})[(2s+2-r)f'(r)+[\frac{2(2s+1)}{(r+2s+1)^2}-\frac{1}{r+2s+1}+V_e(r,s)]]f(r)]=(n-1)f(r)
\end{equation}
gives $V_e(r,s)$ to be
\begin{equation}
V_e(r,s)=\frac{1}{r+2s+1}-\frac{2(2s+1)}{(r+2s+1)^2}
\end{equation}
The added term to eq.(\ref{rad}) takes the form of isospectral potential.
Thus, the new potential is given by $V^+(r,s)=V_o(r,s)+V_e(r,s)$
\begin{eqnarray}
V^+(r,s)=\frac{s(s+1)}{r^2}-\frac{\lambda}{r}+E+\frac{1}{r+2s+1}-\frac{2(2s+1)}{(r+2s+1)^2}
\end{eqnarray}
 It is clear from the above that the relativistic Hydrogen like atom is a conditionally exactly solvable model.
\section{Conclusion}
In this paper, we have constructed the rational extensions to  the Dirac equation in the presence of Hartmann and ring shaped potentials whose solutions are the exceptional Laguerre and exceptional Jacobi polynomials.   We have also constructed the rational extensions to  the  Radial component of the Dirac oscillator wherein the electromagnetic potential is the combination of the Aharnov-Bohm and magnetic monopole potential whose solutions are the exceptional Laguerre  polynomials. Then we have constructed the rational extensions for the relativistic Hydrogen-like atom whose solutions are exceptional Laguerre polynomial.
\section*{Acknowledgments}
KVSSC acknowledges the Department of Science and Technology, Govt of India (fast-track scheme (D. O. No: MTR/2018/001046)) for financial support.
\section*{Appendix1}
calculation of $\frac{dH_n^{(\alpha, \beta)}}{dcos\theta} and \frac{d^2H_n^{(\alpha, \beta)}}{dcos\theta^2}$ 
\begin{eqnarray}
H_n(x)=\frac{f(x)}{x-b}
\end{eqnarray}
\begin{eqnarray}
\frac{dH_n(x)}{dx}=\frac{f'(x)}{x-b}-\frac{f(x)}{(x-b)^2}
\end{eqnarray}
\begin{eqnarray}
\frac{d^2H_n(x)}{dx^2}=\frac{f''(x)}{x-b}-2\frac{f'(x)}{(x-b)^2}+2\frac{f(x)}{(x-b)^3}
\end{eqnarray}
here
$x=cos\theta; b=\frac{2s-1}{2\lambda};$
substituting in the equation(\ref{problem}), we get
\begin{eqnarray}
\\&&\nonumber(1-cos^2\theta)[\frac{f''(cos\theta)}{cos\theta-b}-2\frac{f'(cos\theta)}{(cos\theta-b)^2}+2\frac{f(cos\theta)}{(cos\theta-b)^3}]\\&&\nonumber
+[2(\lambda-scos\theta)-cos\theta][\frac{f'(cos\theta)}{cos\theta-b}-\frac{f(cos\theta)}{(cos\theta-b)^2}]\\&&\nonumber
+[(n^2+2sn)+V_e][\frac{f'(cos\theta)}{cos\theta-b}-\frac{f(cos\theta)}{(cos\theta-b)^2}]
\end{eqnarray}
\begin{eqnarray}
\\&&\nonumber sin^2\theta[f''(cos\theta)-2sin^2\theta\frac{f'(cos\theta)}{(cos\theta-b)}+2sin^2\theta\frac{f(cos\theta)}{(cos\theta-b)^2}]\\&&\nonumber
+[2(\lambda-scos\theta)-cos\theta][\frac{f'(cos\theta)}{cos\theta-b}-\frac{f(cos\theta)}{(cos\theta-b)^2}]\\&&\nonumber
+[(n^2+2sn)+V_e][\frac{f'(cos\theta)}{cos\theta-b}-\frac{f(cos\theta)}{(cos\theta-b)^2}]
\end{eqnarray}
\begin{eqnarray}
\\&& sin^2\theta[f''(cos\theta)\\&&\nonumber
+[2(\lambda-scos\theta)-cos\theta-\frac{2sin^2\theta}{cos\theta-b}]f'(cos\theta)\\&&\nonumber
+[(n^2+2sn)+V_e][\frac{2(\lambda-scos\theta)-cos\theta}{(cos\theta-b)}-\frac{2sin^2\theta}{(cos\theta-b)^2}]f(cos\theta)
\end{eqnarray}
\section*{Appendix2}
Calculation of exceptional potential for the angular part of Hartmann and Ring shaped potentials.
\begin{eqnarray}
\\&&\nonumber sin^2\theta f_n''(cos\theta)\\&&\nonumber +[2\lambda-(2s+1)cos\theta-\frac{2sin^2\theta}{cos\theta-\frac{2s-1}{2\lambda}}]f_n'(cos\theta)\\&&\nonumber
+[n(n+2s)+\frac{2sin^2\theta}{cos\theta-(\frac{2s-1}{2\lambda})^2}
+\frac{2\lambda-2scos\theta-cos\theta}{cos\theta-\frac{2s-1}{2\lambda}}
-\frac{cos\theta}{cos\theta-\frac{2s-1}{2\lambda}}]f_n(cos\theta)= 0
\end{eqnarray}
\begin{eqnarray}
\\&&\nonumber sin^2\theta f_n''(cos\theta)\\&&\nonumber +[\frac{(2\lambda-(2s+1)cos\theta)(cos\theta-\frac{2s-1}{2\lambda})-2sin^2\theta}{cos\theta-\frac{2s-1}{2\lambda}}]f_n'(cos\theta)\\&&\nonumber
+[n^2+2sn+\frac{2sin^2\theta}{cos\theta-(\frac{2s-1}{2\lambda})^2}+\frac{2\lambda-2scos\theta+cos\theta}{cos\theta-\frac{2s-1}{2\lambda}}-\frac{2cos\theta}{cos\theta-\frac{2s-1}{2\lambda}}]f_n(cos\theta) = 0
\end{eqnarray}
\begin{eqnarray}
\\&&\nonumber sin^2\theta f_n''(cos\theta)\\&&\nonumber 
+[\frac{2\lambda cos\theta-2scos^2\theta-cos^2\theta-2s+1-\frac{(2s-1)(2s+1)}{2\lambda}cos\theta-2+2cos^2\theta}{cos\theta-\frac{2s-1}{2\lambda}}]f_n'(cos\theta)\\&&\nonumber
+[n^2+2sn+\frac{2sin^2\theta}{cos\theta-(\frac{2s-1}{2\lambda})^2}+\frac{2\lambda-2scos\theta+cos\theta}{cos\theta-\frac{2s-1}{2\lambda}}-\frac{2cos\theta}{cos\theta-\frac{2s-1}{2\lambda}}]f_n(cos\theta) = 0
\end{eqnarray}
\begin{eqnarray}
\\&&\nonumber sin^2\theta f_n''(cos\theta)\\&&\nonumber +[\frac{2\lambda cos\theta-2scos^2\theta-2s-\frac{(2s-1)(2s+1)}{2\lambda}cos\theta-1+cos^2\theta}{cos\theta-\frac{2s-1}{2\lambda}}]f_n'(cos\theta)\\&&\nonumber
+[n^2+2sn+\frac{2sin^2\theta}{cos\theta-(\frac{2s-1}{2\lambda})^2}+\frac{2\lambda-2scos\theta+cos\theta}{cos\theta-\frac{2s-1}{2\lambda}}-\frac{2cos\theta}{cos\theta-\frac{2s-1}{2\lambda}}]f_n(cos\theta) = 0
\end{eqnarray}
\begin{eqnarray}
\\&&\nonumber sin^2\theta f_n''(cos\theta)\\&&\nonumber 
+[\frac{2\lambda(1-(\frac{2s-1}{2\lambda})cos\theta)cos\theta}{cos\theta-\frac{2s-1}{2\lambda}}-\frac{(2s+1)-\frac{2s-1}{2\lambda}(2s+1)cos\theta}{cos\theta-\frac{2s-1}{2\lambda}}]f_n'(cos\theta)
\\&&\nonumber
+[n^2+2sn+\frac{2sin^2\theta}{cos\theta-(\frac{2s-1}{2\lambda})^2}+\frac{2\lambda-2scos\theta+cos\theta}{cos\theta-\frac{2s-1}{2\lambda}}-\frac{2cos\theta}{cos\theta-\frac{2s-1}{2\lambda}}]f_n(cos\theta) = 0
\end{eqnarray}
\begin{eqnarray}
\\&&\nonumber sin^2\theta f_n''(cos\theta)\\&&\nonumber 
+[\frac{2\lambda(1-(\frac{2s-1}{2\lambda})cos\theta)cos\theta}{cos\theta-\frac{2s-1}{2\lambda}}-\frac{2\lambda[1-(\frac{2s-1}{2\lambda})cos\theta]\frac{2s+1}{2\lambda}}{cos\theta-\frac{2s-1}{2\lambda}}]f_n'(cos\theta)
\\&&\nonumber
+[n^2+2sn+\frac{2sin^2\theta}{(cos\theta-\frac{2s-1}{2\lambda})^2}+\frac{2\lambda(1-(\frac{2s-1}{2\lambda}))cos\theta}{cos\theta-\frac{2s-1}{2\lambda}}-\frac{2cos\theta}{cos\theta-\frac{2s-1}{2\lambda}}]f_n(cos\theta) = 0
\end{eqnarray}
\section*{Bibiliography}

\end{document}